\documentclass[sigconf,authorversion,nonacm]{acmart}
\makeatletter
\def\@copyrightspace{\relax}
\makeatother

\usepackage{diagbox}
\usepackage{graphicx}
\usepackage{newfloat}
\usepackage{listings}
\usepackage{array}
\usepackage{multirow}
\usepackage{csquotes}
\usepackage{fancyvrb}
\usepackage{natbib}
\usepackage{amsmath}
\usepackage{tabularx}
\usepackage[english]{babel}

\AtBeginDocument{%
  \providecommand\BibTeX{{%
    \normalfont B\kern-0.5em{\scshape i\kern-0.25em b}\kern-0.8em\TeX}}}

\usepackage{xcolor}

\newcommand{\gptt}{GPT-3.5}
\newcommand{\gptf}{GPT-4}

\newcommand{\fnolp}{$3{,}028$}
\newcommand{\fnoolp}{$33{,}028$}

\begin{document}

\title[Tracing Journeys of Violent Speech in Incel Posts]{Close to Human-Level Agreement:\\Tracing Journeys of Violent Speech in Incel Posts with \gptf-Enhanced Annotations}


\author{Daniel Matter}
\authornote{Both authors contributed equally to this research.}
\email{daniel.matter@tum.de}
\orcid{0000-0003-4501-5612}
\affiliation{%
  \institution{Technical University of Munich}
  \streetaddress{Richard-Wagner-Str. 1}
  \city{Munich}
  \country{Germany}
  \postcode{}
}

\author{Miriam Schirmer}
\authornotemark[1]
\email{miriam.schirmer@tum.de}
\orcid{0000-0002-6593-3974}
\affiliation{%
  \institution{Technical University of Munich}
  \streetaddress{Richard-Wagner-Str. 1}
  \city{Munich}
  \country{Germany}
  \postcode{}
}

\author{Nir Grinberg}
\email{nirgrn@bgu.ac.il}
\orcid{0000-0002-1277-894X}
\affiliation{%
  \institution{Ben-Gurion University of the Negev}
  \city{Beersheba}
  \country{Israel}
  \postcode{}
}

\author{J{\"u}rgen Pfeffer}
\email{juergen.pfeffer@tum.de}
\orcid{0000-0002-1677-150X}
\affiliation{%
  \institution{Technical University of Munich}
  \streetaddress{Richard-Wagner-Str. 1}
  \city{Munich}
  \country{Germany}
\email{juergen.pfeffer@tum.de}
  \postcode{}
}

\renewcommand{\shortauthors}{}

%
\begin{abstract}
    This study investigates the prevalence of violent language on \emph{incels.is}. It evaluates GPT models (\gptt{} and \gptf{}) for content analysis in social sciences, focusing on the impact of varying prompts and batch sizes on coding quality for the detection of violent speech. We scraped over $6.9M$ posts from \emph{incels.is} and categorized a random sample into non-violent, explicitly violent, and implicitly violent content. Two human coders annotated \fnolp{} posts, which we used to tune and evaluate \gptt{} and \gptf{} models across different prompts and batch sizes regarding coding reliability. The best-performing \gptf{} model annotated an additional $30{,}000$ posts for further analysis.

    Our findings indicate an overall increase in violent speech over time on \emph{incels.is}, both at the community and individual level, particularly among more engaged users. While directed violent language decreases, non-directed violent language increases, and self-harm content shows a decline, especially after 2.5 years of user activity. We find substantial agreement between both human coders ($\kappa = .65$), while the best \gptf{} model yields good agreement with both human coders ($\kappa = 0.54$ for Human A and $\kappa = 0.62$ for Human B). Weighted and macro F1 scores further support this alignment.
    
    Overall, this research provides practical means for accurately identifying violent language at a large scale that can aid content moderation and facilitate next-step research into the causal mechanism and potential mitigations of violent expression and radicalization in communities like \emph{incels.is}.
\end{abstract}


    




\maketitle

\section{Introduction}
The term \enquote{Incels} (\enquote{Involuntary Celibates}) refers to heterosexual men who, despite yearning for sexual and intimate relationships, find themselves unable to engage in such interactions.
The online community of Incels has been subject to increasing attention from both media and academic research, mainly due to its connections to real-world violence \citep{hoffman2020assessing}. Scrutiny intensified after more than 50 individuals' deaths have been linked to Incel-related incidents since 2014 \citep{lindsay2022swallowing}. The rising trend of Incel-related violence underscores societal risks posed by the views propagated within the community, especially those regarding women. In response, various strategic and administrative measures have been implemented. Notably, the social media platform Reddit officially banned the largest Incel subreddit \textit{r/incel} for inciting violence against women \citep{hauser2017reddit}. The Centre for Research and Evidence on Security Threats has emphasized the community's violent misogynistic tendencies, classifying its ideology as extremist \citep{brace2021short}. Similarly, the Texas Department of Public Safety has labeled Incels as an "emerging domestic terrorism threat" \citep{texas2020terrorism}.

Incels mainly congregate on online platforms. Within these forums, discussions frequently revolve around their feelings of inferiority compared to male individuals known as \enquote{Chads,} who are often portrayed as highly attractive and socially successful men who seemingly effortlessly attract romantic partners. Consequently, these forums often serve as outlets for expressing frustration and resentment, usually related to physical attractiveness, societal norms, and women's perceived preferences in partner selection. These discussions serve as an outlet for toxic ideologies and can reinforce patterns of blame and victimization that potentially contribute to a volatile atmosphere \citep{hoffman2020assessing, omalley2022exploration}.

As public attention on Incels has grown, researchers have also begun to study the community more comprehensively, focusing on abusive language within Incel online communities \citep{jaki2019online}, Incels as a political movement \citep{odonnel2022political}, or mental health aspects of Incel community members \citep{broyd2023incels}. 
Despite the widespread public perception that links Incels predominantly with violence, several studies found that topics discussed in Incel online communities cover a broad range of subjects that are not necessarily violence-related, e.g., discussions on high school and college courses and online gaming \citep{mountford2018topic}. 
Nevertheless, the prevalence of abusive and discriminatory language in Incel forums remains a significant concern as it perpetuates a hostile environment that can both isolate members further and potentially escalate into real-world actions.

Although existing research has shed light on essential facets of violence within Incel forums, a comprehensive, computational analysis that classifies various forms of violence expressed in Incel posts remains lacking. Additionally, to the best of our knowledge, no studies focus on trajectories of violent content on a user level.

Understanding violence within the Incel community at the user level is crucial for several reasons. It can provide insights into individual motivations, triggers, and behavioral patterns and reveal the extent of variance within the community, such as what proportion of users engage in violent rhetoric or actions. This nuanced approach could facilitate more targeted and effective intervention and prevention strategies.

\textbf{Scope of this study.} This paper seeks to identify the prevalence of violent content and its evolution over time in the largest Incel forum, \emph{incels.is}. 
We initially perform manual labeling on a subset of the data to establish a baseline and ensure precise categorization for our violence typology. We then employ Open\-AI's \gptt{} and \gptf{} APIs to classify a larger sample of violence identified in online forum threads, thereby enabling a comprehensive annotation of our dataset. We incorporate the human baseline to assess the performance and ensure the accuracy of the categorization process and discuss different experimental setups and challenges associated with annotating Incel posts. 
We then examine how the violent content within the forum evolves for each violence category, looking at the overall share of violent posts within the forum and for individual users within different time frames.

Our main contributions can be summarized as follows:

\begin{itemize}
    \item We find that $15.7\%$ of the posts analyzed in our study ($N =$ \fnoolp{}) exhibit violent speech with a subtle but statistically significant increase over time.
    \item We report a slight decrease in the use of violent language after users have been inactive for a prolonged period.
    \item We perform experiments for annotating data in complex and time-consuming labeling tasks. We present an accessible, resource-efficient, yet accurate state-of-the-art method to enhance data annotation, using manual annotation in combination with \gptf{}.
    \item In particular, we study the effect of batching on the performance of \gptf{} and find that the batch size significantly affects the model's sensitivity.
\end{itemize}

\section{Related Work}

Within computational social science \citep{lazer2009computational}, a diverse body of research has explored the multifaceted landscape of incel posts and forums. 
Natural language processing techniques have been harnessed to analyze the linguistic characteristics of incel discourse, uncovering patterns of extreme negativity, misogyny, and self-victimization. Sentiment analysis, for instance, has illuminated the prevalence of hostile sentiments in these online spaces \citep{jaki2019online, pelzer2021toxic}, while topic modeling has unveiled recurrent themes and narratives driving discussions \citep{baele2021incel, jelodar2021semantic, mountford2018topic}. These studies offer invaluable insights into the dynamics of Incel online communication and serve as a valuable foundation for more comprehensive research to fully understand the complexities of these communities.

\subsection{Incels and Violence}

Due to misogynistic and discriminating attitudes represented in Incel forums, research focusing on violent content constitutes the largest part of academic studies related to this community. \citet{pelzer2021toxic}, for instance, conducted an analysis of toxic language across three major Incel forums, employing a fine-tuned BERT model trained on approximately $20{,}000$ samples from various hate speech and toxic language datasets. Their research identified seven primary targets of toxicity: women, society, incels, self-hatred, ethnicities, forum users, and others. According to their analysis, expressions of animosity towards women emerged as the most prevalent form of toxic language (see \citet{jaki2019online} for a similar approach). On a broader level, \citet{baele2021incel} employed a mix of qualitative and quantitative content analysis to explore the Incel ideology prevalent in an online community linked to recent acts of politically motivated violence. The authors emphasize that this particular community occupies a unique and extreme position within the broader misogynistic movement, featuring elements that not only encourage self-destructive behaviors but also have the potential to incite some members to commit targeted acts of violence against women, romantically successful men, or other societal symbols that represent perceived inequities.

\subsection{Categorizing Violent Language Online}
Effectively approaching harmful language requires a nuanced understanding of the diverse forms it takes online, encompassing elements such as \enquote{abusive language}, \enquote{hate speech}, and \enquote{toxic language} \citep{nobata2016abusive,schmidt2017survey}. Due to their overlapping characteristics and varying degrees of subtlety and intensity, distinguishing between these types of content poses a great challenge. In addressing this complexity, \citet{davidson2017automated} define hate speech as "language that is used to express hatred towards a targeted group or is intended to be derogatory, to humiliate, or to insult the members of the group." Within the research community, this definition is further extended to include direct attacks against individuals or groups based on their race, ethnicity, or sex, which may manifest as offensive and toxic language \citep{salminen2020developing}.

While hate speech has established itself as a comprehensive category to describe harmful language online, the landscape of hateful language phenomena spans a broad spectrum. Current research frequently focuses on specific subfields, e.g., toxic language, resulting in a fragmented picture marked by a diversity of definitions \citep{caselli2020hatebert, waseem2017understanding}. What unites these definitions is their reliance on verbal violence as a fundamental element in characterizing various forms of harmful language. Verbal violence, in this context, encompasses language that is inherently aggressive, demeaning, or derogatory, with the intent to inflict harm or perpetuate discrimination \citep{kansok2023systematic,soral2018exposure,waseem2017understanding}. 
Building on this foundation, we adopt the terminology of \enquote{violent language} as it aptly encapsulates the intrinsic aggressive and harmful nature inherent in such expressions.
To operationalize violent language, \citet{waseem2017understanding} have developed an elaborate categorization of violent language online. This categorization distinguishes between explicit and implicit violence, as well as directed and undirected forms of violence in online contexts and will serve as the fundamental concept guiding the operationalization of violent speech in this paper (see \ref{subsec:violence categories}).
By addressing various degrees of violence, this concept encompasses language employed to offend, threaten, or explicitly indicate an intention to inflict emotional or physical harm upon an individual or group.

\subsection{Classification of Violent Language with Language Models}
Supervised classification algorithms have proven successful in detecting hateful language in online posts. Transformer-based models like HateBERT, designed to find such language, have outperformed general BERT versions in English \citep{caselli2020hatebert}. While HateBERT has proven effective in recognizing hateful language, its adaptability to diverse datasets depends on the compatibility of annotated phenomena. Additionally, although these models exhibit proficiency in discovering broad patterns of hateful language, they are limited in discerning specific layers or categories, such as explicit or implicit forms of violence. The efficiency of the training process is further contingent on the volume of data, introducing potential challenges in terms of time and cost.

Large Language Models (LLMs) present a promising alternative to make data annotation more efficient and accessible. While specialized models like HateBERT often demand significant resources for training and fine-tuning on task-specific datasets, pre-trained LLMs might offer a more flexible, cost-effective solution without requiring additional, expensive transfer learning. 
Recent research has found that using LLMs, particularly OpenAIs GPT variants, to augment small labeled datasets with synthetic data is effective in low-resource settings and for identifying rare classes \citep{moller2023prompt}. Further, \citet{gilardi2023chatgpt} found that \gptt{} outperforms crowd workers over a range of annotation tasks, demonstrating the potential of LLMs to drastically increase the efficiency of text classification.
The efficacy of employing \gptt{} for text annotation, particularly in violent language, has been substantiated, revealing a robust accuracy of $80\%$ compared to crowd workers in identifying harmful language online \citep{li2023hot}. Even in more challenging annotation tasks, like detecting implicit hate speech, \gptt{} demonstrated a commendable accuracy by correctly classifying $80\%$ of the provided samples \citep{huang2023chatgpt}.

While these results showcase the effectiveness of \gptt{} in-text annotation, there remains room for improvement, particularly in evaluating prompts and addressing the inherent challenges associated with establishing a definitive ground truth in complex classification tasks like violent language classification \citep{li2023hot}.


\subsection{User Behaviour in Incel Forums}
The rise of research on the Incel community has also shifted the spotlight on users within the \enquote{Incelverse}, driven by both qualitative and computational approaches. Scholars have embarked on demographic analyses, identifying prevalent characteristics, such as social isolation and prevailing beliefs within the Incelverse. A recent study on user characteristics in Incel forums analyzed users from three major Incel platforms using network analysis and community detection to determine their primary concerns and participation patterns. The findings suggest that users frequently interact with content related to mental health and relationships and show activity in other forums with hateful content \citep{stijelja2023characteristics}. Similarly, \citet{pelzer2021toxic} investigated the spread of toxic language across different incel platforms, revealing that the engagement with toxic language is associated with different subgroups or ideologies within the Incel communities. However, these studies have generally focused on smaller subsets of users and have not examined user behavior across the entirety of the \textit{incels.is} forum. This gap in research is noteworthy, especially when broader studies indicate that content from hateful users tends to spread more quickly and reach a larger audience than non-hateful users \citep{mathew2019spread}. 

\subsection{Summary}
The Incel community has become a subject of growing academic interest due to its complex interplay of extreme views and connections to real-world violence over the last few years. 
While existing studies have shed light on the linguistic and ideological aspects, most have not conducted a thorough user-level analysis across larger forums. 
Our study aims to bridge this research gap by categorizing and examining violent content within \textit{incels.is} itself and at the individual user level. Using manual annotation in conjunction with \gptf{} for this task offers a cost-effective and flexible approach, given its pre-trained capabilities for understanding a wide range of textual nuances. 
\section{Data and Methods}
Besides \emph{incels.is}, platforms like \emph{looksmax.org} and Incel-focused subreddits are key communication channels for the Incel community. After Reddit officially banned the biggest Incel subreddit \textit{r/incel} for inciting violence against women \citep{hauser2017reddit}, many users migrated to alternative platforms. With a self-proclaimed $22{,}000$ members and over 10 million posts\footnote{These numbers are extracted from the landing page and could not be reproduced in our attempts. Out of the $22{,}000$ users, only $11{,}774$ appear to have engaged by posting content.}, \emph{incels.is} has become the leading Incel forum, making it an essential resource for understanding the community.

We scraped all threads from \emph{incels.is}, yielding over $400k$ threads with more than $6.9M$ posts. These were generated by $11{,}774$ distinct users\footnote{This includes $890$ delete users. Once a user deletes their profile, all occurrences of the username are replaced with \emph{Deleted User [XXX]}, but the now anonymous posts are retained.}. We collected the raw HTML responses from the website but ignored all non-text forms of media.

Next, we employed a three-step approach, leveraging the \gptt{}\footnote{\texttt{gpt-3.5-turbo-0301} at temperature $0.1$} and \gptf{}\footnote{\texttt{gpt-4-1106-preview} at temperature $0.1$} APIs. 
Following a round of manual annotation of a random sample of \fnolp{} posts, we iterated prompts and batch sizes for both models to align their classification of violent language within our specified categories with the human baseline.
Finally, we used the best-performing prompt to classify an additional $30{,}000$ posts, which we then analyzed for temporal patterns.

\subsection{Categories of Violence}
\label{subsec:violence categories}

\begin{table}
    \centering
    \caption{Classification examples for each category}
    \label{table:classification_examples}
    \begin{tabular}{m{2.2cm}|m{5.5cm}}
        \hline
        Category & Example \\
        \hline
        \hline
        Non-violent & \textit{Pleasure has become my main purpose of getting new hobbies, music mainly is maintaining me with life.}\vspace{1mm}\\
        \hline
        Explicit, Directed & \textit{I hope the whore gets raped then she can press actual sexual assault charges.}\vspace{1mm}\\
        Explicit, General & \textit{Cliquey, superficial western women deserve the rope, along with the Jews that made them this way.}\vspace{1mm}\\
        Explicit, Self\nobreakdash-Directed & \textit{I'm so ugly I should be killed.}\vspace{1mm}\\
        \hline
        Implicit, Directed & \textit{He looks like he just got back from Auschwitz.}\vspace{1mm}\\
        Implicit, General & \textit{If only women weren't like this. But females love brutality, power, and domination, so in the end they get what they deserve.}\vspace{1mm}\\
        Implicit, Self\nobreakdash-Directed & \textit{The world would be better off without men like me.}\\
    \hline
    \end{tabular}
\end{table}

For categorizing different types of violent language, we used a slightly adapted version of \citet{waseem2017understanding}'s typology of abusive language. To bridge the challenges of navigating through the variety of definitions of hate speech, \citet{waseem2017understanding} have identified mutual characteristics that combine previous classifications of harmful content. This makes their typology a valid reference point when classifying violent language in online forums. 
This concept encompasses expressions that offend, threaten, or insult specific individuals or groups based on attributes such as race, ethnicity, or gender. It further extends to language indicating potential physical or emotional harm directed at these individuals or groups. 
Additionally, differentiating between different types of violence (explicit vs. implicit and general vs. directed) helps gain a more nuanced picture of how violence manifests online.


Following this classification scheme, we distinguish violent posts between explicitly and implicitly violent, as well as between directed, undirected/general, and self-directed violence. Each post is assigned an explicit/implicit and a directed/undirected/self-directed label.
Table \ref{table:classification_examples} provides examples for each category.

In the context of this classification framework, explicit violent language is a very straightforward and usually directly recognizable form of violence, e.g., racist or homophobic threats. While such language can vary in context, it is generally unambiguous in its harmful intent. Implicit violent language is subtler and more challenging to detect. It may involve ambiguous terms or sarcasm and lacks prominent hateful words, making it difficult for human annotators and machine learning algorithms to identify. On the second dimension, directed violent language refers to posts that target a specific individual, either within the forum or outside. General violent language, on the other hand, addresses a group of individuals. In the Incel context, for example, this type of language is often addressed towards women or a specific ethnic group. In our analysis, we focused solely on analyzing the textual content of posts without further differentiating between violent language targeted at particular genders or forum members.

\subsection{Augmented Classification}
Based on this classification scheme, two human annotators independently labeled a subsample of \fnolp{} posts, supported by an annotation manual providing explicit definitions and examples for each violence category. We report Cohen's Kappa ($\kappa$) for intercoder reliability, as it accounts for chance agreement and adjusts for imbalanced data distributions.
We also report weighted and macro F1 scores to assess the performance of the classification against the human baseline.
By involving multiple annotators to establish a human baseline, we ensure a robust assessment of inter-coder consistency, enabling reliable comparisons with the models' annotations.

We used the manually annotated sample of \fnolp{} posts to evaluate the performance of different query prompts and batch sizes for both \gptt{} and \gptf{}. The most relevant queries are presented in the appendix. 
We started with a very basic prompt that only included information on our classification scheme, i.e., the categories of violence.
Adding contextual information, specifically about the posts originating from an Incel forum, significantly improved the model's performance.
To keep iterating on the prompt, we kept looking at posts where the model's classification differed from the manual annotation and tried to find patterns in the misclassifications. Further, we used a form of self-instruction, presenting those misclassifications to the model itself and asking it for advice on improving the prompt.

\gptt{} allows for a maximum of $4k$ tokens for input and output, which can contain multiple messages with different roles. \gptf{} has a context window of $128k$ tokens.
We provided the instruction part of the prompt as a \emph{system} message, while the posts were delivered in a second \emph{user} message.
This allowed us to batch multiple posts into a single classification request, making the speed and cost of the classification process manageable. Without batching, reiterating the same prompt for each post would substantially inflate the required number of tokens. 
We experimented with different batch sizes, ranging from $10$ to $200$ posts per batch.

In practice, each classification batch looked like
\begin{Verbatim}[samepage=true]
[System Message] 
<Prompt>
The posts are:
\end{Verbatim}
followed by the batch of posts

\begin{Verbatim}[samepage=true]
[User Message]
Post 1: <Post 1>
Post 2: <Post 2>
...
\end{Verbatim}

\gptf{} introduces a novel JSON output mode, enabling the model to generate outputs in a JSON object format instead of plain text. This output format must be specified within the prompt. Our findings indicate that this mode does not alter the model's performance but significantly simplifies the process of parsing its outputs. For all our final classifications, this mode was utilized.

Regarding data preprocessing, we limited our intervention to consolidating multiple new lines into one line. We found the model could handle the posts' raw text very well. Notably, it did not miss or confuse any post at any time.
After iterating over the queries, we chose the one that performed best against the human baseline to annotate another $30{,}000$ posts.

\subsection{Time-Based Patterns of Violent User Posts}

After obtaining the final set of \fnoolp{} annotated posts, we matched them across users and time to perform a time-based analysis of the prevalence of violent language. Each user was assigned a timeline to determine when they were active, for how long, and when they were inactive. 
We repeated this analysis multiple times, considering different timespans between posts as a period of activity or inactivity to examine the impacts of activity within the forum.

We consider timespans of one hour, six hours, 12 hours, one day, two weeks, and six months (180 days). 
If a user was inactive for at least the respective timespan, i.e., they did not post within that timespan, we consider them inactive. 
If consecutive posts are less than the respective timespan apart, we consider them part of the same session.

We perform an ordinary least squares linear regression on each of those timespans, using the share of violent posts as the dependent variable and the time since the user's first post of the session as the independent variable.
We focus on the coefficient ($\beta$), providing both its numerical values and its associated level of statistical significance ($*$ indicating $p < 0.05$, ${*}{*}$ indicating $p < 0.01$, and ${*}{*}{*}$ indicating $p < 0.001$) in the results section.

\section{Results}

\subsection{Performance of Automated Classification}

\begin{table*}[ht]
	\centering
	\renewcommand{\arraystretch}{1.1}
	\begin{tabularx}{\textwidth}{lXXXXXXXXX}
		\toprule
			 & Human A & Human B & GPT3.5 & GPT4/10 & GPT4/20 & GPT4/50 & GPT4/100 & GPT4/200\\
			\hline
			Human A & - & .69/.85/.77 & .40/.70/.52 & .53/.74/.63 & \textbf{.54*}/\textbf{.76*}/\textbf{.63*} & .52/.74/.62 & .52/.75/.60 & .36/.71/.49 \\
			Human B & .69/.87/.77 & - & .39/.75/.54 & .58/.79/.67 & .55/.79/.65 & .61/.83/\textbf{.67*} & \textbf{.62*}/\textbf{.84*}/.67 & .40/.77/.52 \\
			GPT3.5 & .40/.67/.52 & .39/.68/.54 & - & \textbf{.54*}/\textbf{.75*}/\textbf{.62*} & .49/.72/.59 & .49/.71/.59 & .47/.70/.56 & .37/.67/.48 \\
			GPT4/10 & .53/.73/.63 & .58/.76/.67 & .54/.74/.62 & - & \textbf{.75*}/\textbf{.86*}/\textbf{.78*} & .60/.77/.67 & .58/.76/.66 & .46/.68/.55 \\
			GPT4/20 & .54/.75/.63 & .55/.77/.65 & .49/.74/.59 & \textbf{.75*}/\textbf{.87*}/\textbf{.78*} & - & .69/.83/.74 & .65/.81/.71 & .44/.71/.51 \\
			GPT4/50 & .52/.77/.62 & .61/.82/.67 & .49/.76/.59 & .60/.80/.67 & .69/.85/\textbf{.74*} & - & \textbf{.72*}/\textbf{.87*}/.72 & .47/.75/.55 \\
			GPT4/100 & .52/.78/.60 & .62/.84/.67 & .47/.77/.56 & .58/.80/.66 & .65/.84/.71 & \textbf{.72*}/\textbf{.88*}/\textbf{.72*} & - & .51/.80/.59 \\
			GPT4/200 & .36/.77/.49 & .40/.81/.52 & .37/.79/.48 & .46/.79/.55 & .44/.80/.51 & .47/.82/.55 & \textbf{.51*}/\textbf{.83*}/\textbf{.59*} & - \\
		\bottomrule \\ 
	\end{tabularx}
	\caption{Cohen's Kappa / Weighted F1-Score / Macro F1-Score. Asterix indicates the best performance per row, excluding humans. For the F1-scores, left indicates the ground-truth, while top indicates predictions.}
	\label{tab:agreements}
\end{table*}
\begin{table}[ht]
	\centering
	\renewcommand{\arraystretch}{1.1}
	\begin{tabularx}{0.475\textwidth}{lXXXXXXX}
		\toprule
			 & s=10 & s=20 & s=50 & s=100 & s=200 & H\nobreakdash-$\varnothing$\\
			\hline
			Non. & 0.58 (1.00) & 0.62 (1.07) & 0.70 (1.20) & 0.72 (1.24) & 0.82 (1.41) & 0.70 (1.21) \\
			Expl. & 0.28 (1.00) & 0.26 (0.96) & 0.21 (0.78) & 0.22 (0.80) & 0.16 (0.57) & 0.22 (0.81) \\
			Impl. & 0.14 (1.00) & 0.12 (0.81) & 0.09 (0.61) & 0.06 (0.41) & 0.02 (0.16) & 0.07 (0.52) \\
		\bottomrule \\ 
	\end{tabularx}
	\caption{Class distribution for different batch sizes $s$.}
	\label{tab:batch}
\end{table}

Table \ref{tab:agreements} shows the pairwise Cohen's Kappa and weighted/macro F1 scores of all relevant annotation methods. 
Human A and B indicate the two human annotators, while \gptt{} presents the best-performing \gptt{} query and batch-size combination.
\gptf{}/X showcases the performance of \gptf{} with batch-size $X$ for the best-performing query, each.

All best queries asked the models to provide reasons for their decision, usually in the form of the most important words.
Models must give these reasons before being asked to classify the posts, as they cannot respond to reasons for their decision after the fact.

We decided to include Cohen's Kappa as a measure of inter-annotator agreement, as it is commonly used in social sciences.
The weighted F1 score differentiates between ground truth and predicted labels, making it a suitable metric for comparing the performance of the models against the human annotators. The macro F1 score, on the other hand, is a suitable metric for inspecting the performance regarding underrepresented classes, as it computes the F1 score for each class individually and then takes the average of those scores.

\gptt{} is outperformed by \gptf{} in all metrics when comparing its labels against both human annotators. The rest of the analysis hence focuses on the performance of the different \gptf{} variants.

The inter-annotator agreement between Human A and Human B, as measured by Cohen's Kappa ($\kappa$), is $0.69$, indicating a substantial level of agreement. Their weighted and macro F1 scores of $0.85$ and $0.77$, respectively, illustrate apt performance with distinct yet varying levels of precision and recall in their annotations.
Overall, Human A is more likely to label a post as violent than Human B, with $66\%$ of posts labeled as violent by Human A, compared to $75\%$ by Human B.

The analysis of different batch sizes reveals notable variations in the performance of \gptf{}. Batch size $20$ shows the highest agreement with Human A, as evidenced by its superior performance metrics. 
Conversely, batch size $100$ aligns more closely with Human B, particularly regarding $\kappa$ and weighted F1 scores. For the macro F1 score, batch size $50$ exhibits the best alignment with Human B. 
The achieved Kappa values of $0.54$ against Human A and $0.62$ against Human B indicate moderate to substantial agreement.
Macro and weighted F1 scores of $0.63$ and $0.76$ against Human A and $0.67$ and $0.84$ against Human B, respectively, indicate a high level of precision and recall in the classification of all three categories.

Table \ref{tab:batch} elucidates the overall label distribution across varying batch sizes, in which we observe a statistically significant shift. With increasing batch sizes, there is a discernible trend of fewer posts being classified as explicitly or implicitly violent and more as non-violent. 
This trend is more pronounced in the classification of implicit violence. Using a batch size of $10$, $14\%$ of all posts were labeled as implicitly violent. At batch size $200$, this drops by $84\%$ to $2\%$ of the total posts. The share of posts labeled as explicitly violent only decreases by $43\%$ from $28\%$ to $16\%$.

The label distribution generated at batch size $50$ most closely aligns with the average distribution generated by the human annotators, suggesting an optimal batch size for achieving a human-like understanding in content classification.

We further investigated the correlation between a post's position in a batch and its likelihood of being labeled as violent. 
Posts positioned later in the batch were less frequently tagged as violent for larger batch sizes.
This trend was consistent across different batch sizes but did not reach statistical significance.

Due to the high level of agreement with humans A and B and the match in the overall class distribution, we used the labels generated by \gptf{} with batch size $50$ for the remainder of our analysis.

\subsection{Time-Based Patterns of Violent User Posts}

Our results indicate that posts containing violent language, whether explicit or implicit, constitute $\sim 15\%$ of all posts. 
Explicitly violent language accounts for $\sim 9\%$, while implicitly violent language accounts for $\sim 6\%$. This leaves $\sim 85\%$ of forum posts non-violent. 

The user analysis reveals a wide range of engagement levels, with an average of about $586$ posts and a median of $24$ posts per user. About $10\%$ of users maintained forum activity for at least $2.5$ years at the time of scraping, highlighting their sustained engagement.
Approximately $23.8\%$ of forum users contributed only one post, underscoring the presence of occasional contributors within the platform's user community, while the $10\%$ most active users have posted at least $1152$ times.

\begin{figure}
    \centering
    \includegraphics[width=0.47\textwidth]{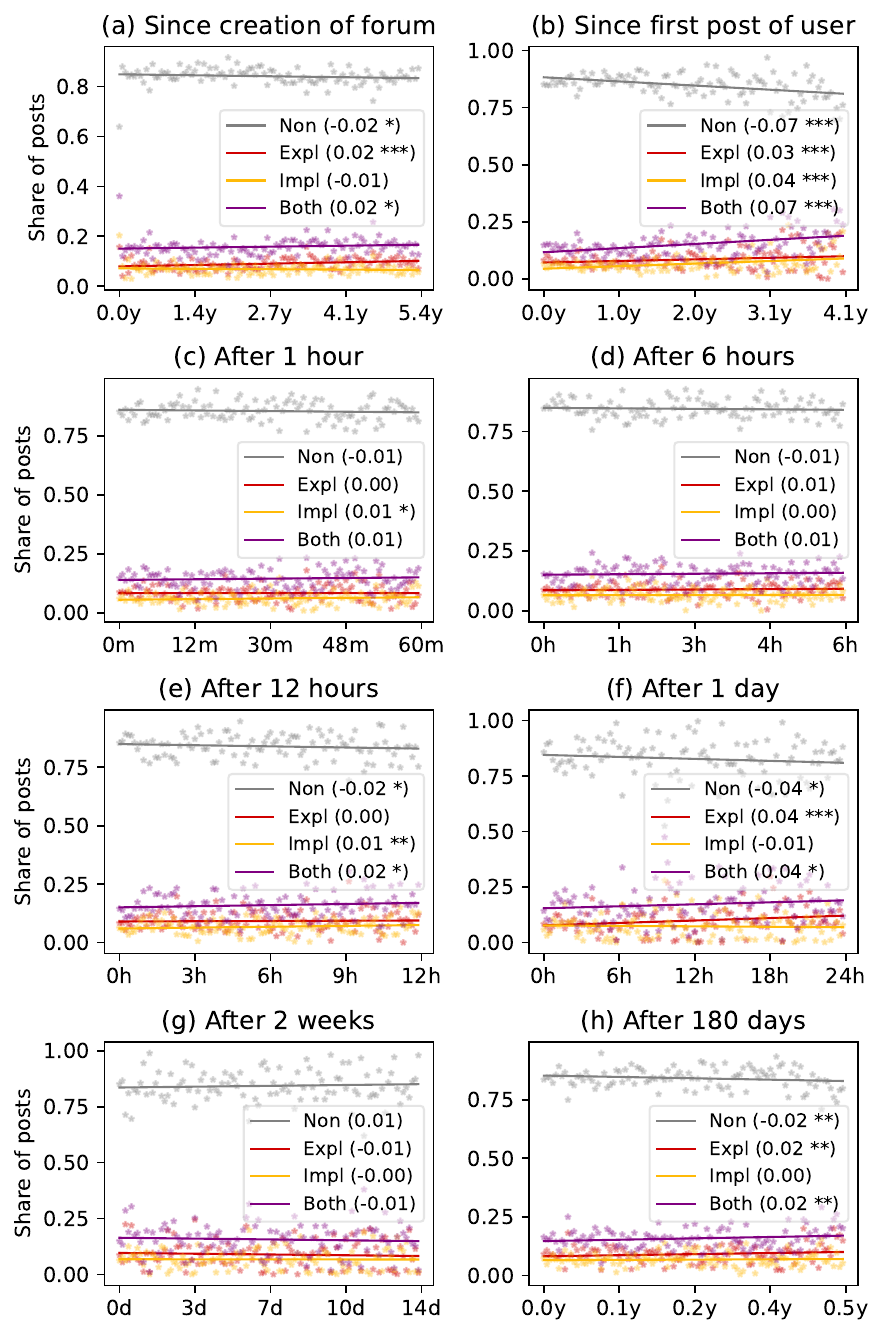}
    \caption{Linear Regression between time and share of violent posts. Numbers indicate the slope of the regression line, and stars indicate statistical significance.}
    \label{fig:timeseries}
\end{figure}

Figure \ref{fig:timeseries} (a-h) illustrates the temporal evolution of violent language in posts, with different time intervals as predictors for the share of posts in each violence category. Our results indicate that since the forum's creation (a) about five years ago, violent language overall has been slightly increasing on a statistically significant level ($\beta = 0.02$ for the combined violence category), accompanied by a concurrent decline in non-violent language ($\beta = -0.02$). 
This shift is primarily driven by an increase in explicit violence, with implicit violence decreasing over time.

Following individual user journeys (b), we find that posts become more violent ($\beta = 0.07$ for the combined violence category) over multiple years.
Most of this change happens after a tipping point at around one year in the forum, before which no statistically significant change in violent language is observable.
The incline in violent language is distributed over both explicit and implicit violence, with the latter being more pronounced ($\beta = 0.04$ for implicit violence, $\beta = 0.03$ for explicit violence). 
The share of non-violent language accordingly decreases over time ($\beta = -0.07$).

Subfigures (c-h) explore the impact of temporary inactivity on the prevalence of violent language.
Each figure follows users for a period p, as indicated in the respective subfigure. The tracking takes place after these specific users have remained inactive for at least the same designated period.

From these figures, we observe varying results. For both one-hour (c), as well as six-hour (d) intervals, no statistically significant change in violent language is observable.
Within the 12-hour (e), as well as the one-day interval (f), we observe a slight increase in violent language overall ($\beta = 0.02$ and $\beta = 0.04$, respectively), accompanied by a decrease in non-violent language ($\beta = -0.02$ and $\beta = -0.04$, respectively) over those respective periods, all of which are weakly statistically significant.
This indicates that users appear to experience a \emph{cool-off} effect. Once they have been inactive for at least 12 hours, and more strongly if they have been inactive for at least 24 hours, they are less likely to post violent language than at the end of an active session of the same length.

However, this effect only holds for meso-timescales. The two-week interval reveals no significant change, while the six-month interval (h) already starts to slowly pick up on the overall trend towards more violent language ($\beta = 0.02$), which has already been observed on the macro-timescale (b).

\begin{figure}
    \centering
    \includegraphics[width=0.47\textwidth]{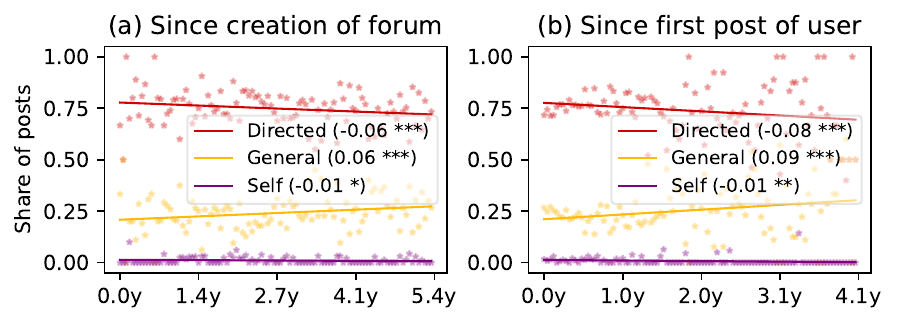}
    \caption{Linear Regression between time and category of directedness. Numbers indicate the slope of the regression line, and stars indicate statistical significance.}
    \label{fig:timeseries:dir}
\end{figure}

Figure \ref{fig:timeseries:dir} showcases the same analysis for the different categories of directedness. Subfigures (c-h) are omitted since they do not contain any statistically relevant results, indicating that no substantial change in directed, general, or self-directed violence can be observed within the examined time frames. 
Subfigures (a) and (b) reveal, that the share of directed violence is decreasing over time, both for the forum, as well as for individual users. 
This is accompanied by an increase in non-directed (general) violence, which is more pronounced for individual users ($\beta = 0.09$) than for the forum as a whole ($\beta = 0.06$).
The share of self-harm content makes up $\sim 1.1 \%$ of all posts and is decreasing over time for both the forum and individual users (both $\beta = -0.01$). Interestingly, in individual user journeys (b), after $\sim 2.5$ years of activity the absolute number of posts marked as self-harm drops to zero. 

\section{Discussion and Limitations}

The difficulty in detecting certain kinds of violent language differs significantly between categories.
While explicit acts of violence, such as physical assault or overt verbal abuse, may be easier to detect through keywords or contextual cues, implicit violence often manifests in more nuanced ways that are hard even for humans to identify \citep{strathern2022identifying}. These include coded language that carries a threatening subtext. For instance, users often refer to Elliot Rodger, who committed an Incel-related attack in 2014, stating posts like \enquote{Just go ER.} Also, Incel-specific language is frequently inherently derogative towards women, calling them \emph{foids}, short for feminine humanoids, and uses racist slang, e.g., \emph{Currycel} for an Indian Incel.
Herein lies an apparent strength of LLMs, which proved to be very effective at finding and classifying these Incel-specific terms. Having been trained on large parts of the internet, it is very probable that the model has encountered these terms before and learned to associate them with violence.



Looking at the classification framework on a content level, while \citet{waseem2017understanding}'s typology provides a reasonable foundation, an Incel-adapted framework could yield more accurate results. For instance, \citet{pelzer2021toxic} outline a framework that differentiates between posts targeting women, society, Incels themselves, ethnic groups, forum users, self-hatred, and other categories. For future studies, it might be interesting to incorporate these target references to examine whether violent content varies in that regard. This seems especially relevant given that a substantial amount of violent posts is not directed to women but also to "Chads," "normies" (referring to average, ordinary individuals in Incel terms), and society in general.

\subsection{Classifying Violent Language with GPT}
With an attained intercoder agreement of $\kappa = 0.54$ and $\kappa = 0.62$ with the human baselines, \gptf{} demonstrates moderate to substantial performance in annotating various categories of violence in Incel posts. 
Given an appropriate query and batch size, the model reproduces the overall label distribution and achieves reasonable performance within each class, as indicated by macro F1 scores of $0.63$ and $0.67$.
Our weighted F1 scores of $\sim 0.8$ align with those reported by other studies, including \citet{huang2023chatgpt} and \citet{li2023hot}.


Our study indicates that LLMs can produce a sensible starting point for the zero-shot classification of violent content, providing a solid foundation for further analyses. 
When asking the model to explain its decision, we found its explanation constitutes a valuable point of reference for human annotators.

We did not evaluate its capabilities on a standardized corpus. Other models, such as HateBERT \citep{caselli2020hatebert}, may perform better on datasets they are fine-tuned on. At the same time, it is noteworthy that even hate-speech-specific models encounter challenges when categorizing different types of violent content, highlighting the intricate nature of the task \citep{poletto2021resources,yin2021towards}. Additionally, these models may not be explicitly designed to differentiate within distinct categories of violent language, introducing an additional layer of complexity to the classification process.

While the change in sensitivity for different batch sizes might seem discerning at first, it can also be understood as a tuneable hyperparameter.
During query optimization, we found that manipulating the model's overall sensitivity by altering the query, as opposed to sensitivity towards a specific class, is challenging. The batch size provided us with a valuable tool to adjust the overall sensitivity, allowing us to match the overall label distribution to fit between that of the human annotators.
It is worth noting that this adjustment substantially impacts the model's speed and cost, as discussed in Section \ref{sec:cost}.

The substantial agreement with the human annotators, combined with its accessibility and cost-effectiveness, render \gptf{} as a reasonable alternative to traditional, embedding-based classification models. 
Moreover, employing LLMs to augment the annotated sample offers distinct advantages, as it spares human annotators from the potential emotional distress of reading content containing violence against specific individuals or groups.

\subsection{Users and Violence over Time}
\textbf{General Posting Behaviour.} 
The analysis of user engagement within the online forum reveals notable variations in user behavior. While an average of $586$ posts per user appears substantial, a median of $24$ posts per user indicates a very skewed distribution. 
This pattern is even more pronounced with roughly $10\%$ of users showing sustained engagement spanning at least $2.5$ years. In contrast, nearly a quarter of the user population comprises occasional contributors characterized by only having posted once. These findings underscore the diverse spectrum of user activity within the platform, ranging from highly engaged, long-term participants to sporadic contributors with limited involvement.

\textbf{General Patterns Over Time.} 
Our analysis reveals nuanced patterns in the temporal evolution of violent language within the forum. Over the five years since the forum's creation, we identified a statistically significant but modest increase in violent content, accompanied by a concurrent decline in non-violent content. Notably, this trend primarily results from an increase in explicit violence. In this context, it is essential to note that implicit violence exhibits higher variance compared to explicit violence, making the latter more probable to show statistically significant effects. 
This underscores the inherent challenge of classifying implicit violence, emphasizing the importance of validation through qualitative analysis. It also emphasizes the need to define clearer parameters for computational analysis to make implicit violence more tractable from a computational perspective.

When examining individual user behavior, posts become progressively more violent over time. 
On an individual level, this suggests that users tend to contribute increasingly violent content as their engagement duration extends. 
When combined, these findings highlight a twofold trend: an overall, forum-wide increase in violence, as well as a more pronounced shift towards violent language in contributions from individual users over extended periods of engagement.
These trends may be explainable by evolving community norms, which become more tolerant towards violent content over time, user familiarity, or moderation effects \citep{gibson2019free}.

Looking at the evolution of violent posts regarding their directedness, we observe a declining trend in directed violence and an increasing trend in non-directed (general) violence, both at the forum level and among individual users. This shift implies a change in the type of aggression within the community, where users resort to more generalized hostility. 
Understanding the driving factors behind this increase is essential to address and mitigate the overall aggression levels within the forum. It might prove fruitful to examine whether the generalized violence is directed towards a specific target group (e.g., women, non-Incel men, etc.) \citep{pelzer2021toxic}.
At the same time, self-directed violent content slightly decreased in the overall forum and completely dropped to zero on an individual user level after three years. Reasons for this may include psychosocial factors, such as peer support or individual psychological characteristics \citep{broyd2023incels,speckhard2022self}.

\textbf{Patterns Within Specific Time Intervals.}
The analysis of violent language development across different time intervals reveals relative stability in the prevalence of violent language within the short-term periods, i.e., the one-hour and six-hour time windows. User engagement with the platform within these brief durations appears not to change violent content creation.
The most notable trend occurs within the one-day timeframe, where we observe a substantially lower level of violent language at the beginning of the interval, indicating \enquote{cool-off effect} after users have been inactive for at least 24 hours.

Although not statistically significant, the two-week period stands out from the observed patterns, as it exhibits a decrease in violent language over time. While this might initially appear as an anomaly, this deviation could result from chance or other factors not accounted for in the current analysis. Therefore, it might be valuable to validate these findings with additional data to determine the reliability of this particular observation.

\textbf{Escalating Violence within the Incel Community and Beyond.} 
The results of our study align with previous research focused on radicalization within the Incel community. As noted by \citet{habib2022making}, users who become part of online Incel communities exhibit a $24\%$ increase in submitting toxic content online and a $19\%$ increase in the use of angry language. The authors conclude that Incel communities have evolved into platforms that emphasize expressing anger and hatred, particularly towards women. In the context of online discussions on conspiracy theories, \citet{phadke2022pathways} modeled various radicalization phases for Reddit users, identifying different stages in radicalization, that could also be applied to the Incel context in future studies. 

Additionally, individual beliefs and attitudes of users, such as their identification with the \enquote{red pill} or the more extreme \enquote{black pill}, could correlate with the observed trends. It is plausible that belonging to a particular ideological subgroup, such as black-pill adherents, may influence how members express violent content. These ideologies may affect the time spent online, the duration of active online engagement, and the posting frequency, making them relevant factors to consider in this context. 
It might be fruitful to examine whether the observed trends are more pronounced among specific subgroups within the community or whether they are evenly distributed over the user population.

Although our results are too subtle to account for an actual pattern of radicalization, it might also be interesting to build upon these results and dive more deeply into the content of violent posts within specific time windows to see if phases of escalation can be identified.

\subsection{GPT Cost and Speed}
\label{sec:cost}

For the scope of this study, we spend a total of $\sim \$50$ for OpenAI's APIs. 
This includes many iterations over all the human-annotated posts, as well as the additionally annotated posts, and a lot of trial and error.
Overall, we estimate \gptt{} and \gptf{} annotated $\sim 100{,}000$ posts, which amounts to $\sim \$ 0.0005$ per annotated post.

A key component of keeping the cost low is proper input batching.
Our prompts are around $500$ tokens long, whereas the average post is around $50$ tokens long.
Naively sending each post individually would have cost $550T \times \frac{\$0.01}{1000T} = \$ 0.0055$ per post, or $\$180$ for the final set of \fnoolp{} annotated posts.
Increasing the batch-size to $50$ yields a cost per batch of $3000T \times \frac{\$0.01}{1000T} = \$ 0.03$, or $\$20$ for the final set of \fnoolp{} annotated posts.
\gptt{} is significantly cheaper.

The average time for \gptf{} to annotate a single post was $1$ second at batch size $50$. 
The total time for \gptf{} to annotate $100{,}000$ posts was $\sim 28$ hours.
On multiple occasions, we experienced significant slow-downs in the APIs' response time, which are confirmed by OpenAI\footnote{\url{https://status.openai.com}}.
Moving our long-running jobs to the early European morning significantly improved the experience of working with the API.

\subsection{Summary and Future Work} 
Our study reveals a subtle but statistically significant increase in overall violence on \emph{incels.is} within the forum. The same trend is found to be more pronounced on the user level, where the prevalence of violent content strongly increases over multiple years of activity. 
Additionally, directed violence decreases over time, yielding an increase in non-directed violence, while self-harm content gradually diminishes within the forum and among individual users. 

Overall, these findings highlight the complex relationship between user engagement duration and violent content generation. Further research may be needed to explore the underlying motivations and dynamics driving these temporal patterns in online Incel discussions. Exploring broader time-related factors, including the potential impact of COVID-19-related dynamics on online behavior, is particularly pertinent, given prior research indicating shifts in behavior patterns during the pandemic that contributed to heightened radicalization in various online forums, including those within Incel communities \citep{davies2021witch}. Additional (computational) studies and in-person surveys with community members could provide deeper insights and guide interventions to foster more positive interactions within the forum.


Our study emphasizes the effectiveness of leveraging LLMs, specifically \gptf{}, as annotators in intricate classification tasks, especially in identifying different types of violent content in online communities — an inherently challenging task for human annotators. Notably, we find a significant agreement between \gptf{} and both human annotators, providing strong support for its practicality in these applications.
Additionally, by providing reasons for its classification, \gptf{} can drastically streamline situations where human annotators are uncertain. While our results provide a baseline, further research is needed to evaluate the performance of \gptf{} compared to other hate-speech-focused models.

\section{Ethical Statement}
This paper follows the ACM Code of Ethics and Professional Conduct. Our data processing procedures did not involve any handling of private information. 
The user names obtained during the scraping process do not contain sufficient and valid information to make conclusions about online users' personal information. The same is true for posts directly cited in this paper. 
Despite offensive language in these posts, we have included them to enhance the clarity and understanding of our categorization for our readers. To promote responsible usage and prevent potential misuse, we make our annotated data available upon request, considering the data's violent nature and the need to address concerns about circumventing hate speech detection.
Both human annotators were informed of and aware of the potentially violent content in Incel posts before the annotation process, with the ability to decline annotation at any time. Both coders were given the chance to discuss any distressing material encountered during annotation. As discussions on the potential trauma or adverse effects experienced by annotators while dealing with hate speech become more prevalent \citep{kennedy2022introducing}, we have proactively provided annotators with a recommended written guide designed to aid in identifying changes in cognition and minimizing emotional risks associated with the annotation process.

\bibliographystyle{ACM-Reference-Format}
\bibliography{sources}

\appendix
\section{Appendix}
\textbf{Incel Hint}
{\small You are the moderator of an online incel forum, trying to mitigate abusive, hateful language. 
You are given posts from the forum and need to label them. 
Each post can be non-violent (N), or contain explicit violence / hate (E) or implicit violence / hate (I). 
If a post contains violence or abusive language, it can either be directed at a specific person (D), be undirected (U), or be self-directed at the author of the post (S).
As it is an incel forum, please pay special attention to hate against women, and any incel-specific language.
Explicit violence is a very straightforward and usually directly recognizable form of violence. Implicit abuse may involve ambiguous terms or sarcasm and lacks obvious hateful words.
Directed violent content targets a specific individual, either within the forum or outside. 
Undirected violence is aimed at groups of people, rather than individuals.
Please pay special attention to derogatory and dehumanizing incel slang. In particular misogynistic and offensive language, as well as acronyms such as foid, normfags, and ER are commonly used in both explicitly, as well as implicitly violent language.
Some messages contain citations of other posts. Please only consider the newest message, i.e., the part after "click to expand".}

\noindent
\textbf{Give Reason}
{\small You are the moderator of an online incel forum, trying to mitigate abusive, hateful language. 
You are given posts from the forum and need to label them. 
Each post can be non-violent (N), or contain explicit violence / hate (E) or implicit violence / hate (I). 
If a post contains violence or abusive language, it can either be directed at a specific person (D), be undirected (U), or be self-directed at the author of the post (S).
As it is an incel forum, please pay special attention to hate against women, and any incel-specific language.
Explicit violence is a very straightforward and usually directly recognizable form of violence.
Implicit abuse may involve ambiguous terms or sarcasm and lacks obvious hateful words.
Directed violent content targets a specific individual, either within the forum or outside. 
Undirected violence is aimed at groups of people, rather than individuals.
Please pay special attention to derogatory and dehumanizing incel slang. In particular misogynistic and offensive language, as well as acronyms such as foid, normfags, and ER are commonly used in both explicitly, as well as implicitly violent language.
Some messages contain citations of other posts. Please only consider the newest message, i.e., the part after "click to expand".
To respond, return the number of the post and the classification letter.
If you detect any kind of violence, also give the two to three most important words from the message to detect that violence.}


\noindent
\textbf{Give Reason, Few Examples}
{\small You are the moderator of an online incel forum, trying to mitigate abusive, hateful language. 
You are given posts from the forum and need to label them. 
Each post can be non-violent (N), or contain explicit violence / hate (E) or implicit violence / hate (I). 
If a post contains violence or abusive language, it can either be directed at a specific person (D), be undirected (U), or be self-directed at the author of the post (S).
As it is an incel forum, please pay special attention to hate against women, and any incel-specific language.
Explicit violence is a very straightforward and usually directly recognizable form of violence.
Implicit abuse may involve ambiguous terms or sarcasm and lacks obvious hateful words.
Directed violent content targets a specific individual, either within the forum or outside. 
Undirected violence is aimed at groups of people, rather than individuals.
Please pay special attention to derogatory and dehumanizing incel slang. In particular misogynistic and offensive language, as well as acronyms such as foid, normfags, and ER are commonly used in both explicitly, as well as implicitly violent language.
Only consider proposed violence. The mere mention or informative description of violent events is not relevant to your classification.
Posts should only be marked as violent, if they endorse a past act of violence, or propose a future act of violence. 
Further, discriminatory language is also considered violent.
Some messages contain citations of other posts. Please only consider the newest message, i.e., the part after "click to expand".
To respond, return the number of the post and the classification letter.
If you detect any kind of violence, give a brief description of the violence before your classification.
Here are some examples:
Post 1: 
Raping a blind woman should be easy
Post 2:
If only women weren't like this. Then things like height, frame, status, and dark triad traits would matter less to them. But females love brutality, power, and domination. Where modern soyciety went wrong was trying to put women (who are naturally submissive) into positions equal to or above men.
Post 3:
Nature doesn't give a fuck about us and yet we are expected to wageslave like good little boys.}

\noindent
\textbf{Final Prompt}
{\small You are the moderator of an online forum, trying to mitigate abusive, hateful language. 
You are given posts from the forum and need to label them. 
Each post can be non-violent (N), or contain explicit violence / hate (E) or implicit violence / hate (I). 
If a post contains violence or abusive language, it can either be directed at a specific person (D), be undirected (U), or be self-directed at the author of the post (S).
As it is an incel forum, please pay special attention to hate against women, and any incel-specific language.
Explicit violence is a very straightforward and usually directly recognizable form of violence.
Implicit abuse may involve ambiguous terms or sarcasm and lacks obvious hateful words.
Directed violent content targets a specific individual, either within the forum or outside. 
Undirected violence is aimed at groups of people, rather than individuals.
An example of explicit, directed violence would be: the stupid foid should be punished.
An example of implicit, undirect violence would be: I would rather date apes than women.
For every post, justify your classification no more than 5 words.
Skip all non-violent posts.
Humans have a lower threshold for labeling something as violence, in particular implicit violence.
Please be very sensitive to these forms of violence and label posts carefully.}


\end{document}